\documentclass[twocolumn]{aastex61}
\usepackage{amsmath,amstext}
\usepackage[T1]{fontenc}
\usepackage{apjfonts} 
\usepackage[figure,figure*]{hypcap}
\usepackage{breakurl}

\shorttitle{cat-puma}
\shortauthors{Jiajia Liu et al.}

\begin{document}

\title{A New Tool for CME Arrival Time Prediction Using Machine Learning Algorithms: CAT-PUMA}

\author{Jiajia Liu}
\email{jj.liu@sheffield.ac.uk}
\affiliation{Solar Physics and Space Plasma Research Center (SP2RC), School of Mathematics and Statistics, The University of Sheffield, Sheffield S3 7RH, UK}

\author{Yudong Ye}
\affiliation{SIGMA Weather Group, State Key Laboratory of Space Weather, National Space Science Center, Chinese Academy of Sciences, Beijing 100190, China}
\affiliation{University of Chinese Academy of Sciences, Beijing 100049, China}

\author{Chenglong Shen}
\affiliation{CAS Key Laboratory of Geospace Environment, Department of Geophysics and Planetary Sciences, University of Science and Technology of China, Hefei, Anhui 230026, China}
\affiliation{Synergetic Innovation Center of Quantum Information \& Quantum Physics, University of Science and Technology of China, Hefei, Anhui 230026, China}

\author{Yuming Wang}
\affiliation{CAS Key Laboratory of Geospace Environment, Department of Geophysics and Planetary Sciences, University of Science and Technology of China, Hefei, Anhui 230026, China}
\affiliation{Synergetic Innovation Center of Quantum Information \& Quantum Physics, University of Science and Technology of China, Hefei, Anhui 230026, China}

\author{Robert Erd{\'e}lyi}
\affiliation{Solar Physics and Space Plasma Research Center (SP2RC), School of Mathematics and Statistics, The University of Sheffield, Sheffield S3 7RH, UK}
\affiliation{Department of Astronomy, E\"{o}tv\"{o}s Lor\'{a}nd University, Budapest, P\'{a}zm\'{a}ny P. s\'{e}t\'{a}ny 1/A, H-1117, Hungary}

\begin{abstract}
Coronal Mass Ejections (CMEs) are arguably the most violent eruptions in the Solar System. CMEs can cause severe disturbances in the interplanetary space and even affect human activities in many respects, causing damages to infrastructure and losses of revenue. Fast and accurate prediction of CME arrival time is then vital to minimize the disruption CMEs may cause when interacting with geospace. In this paper, we propose a new approach for partial-/full-halo CME Arrival Time Prediction Using Machine learning Algorithms (CAT-PUMA). Via detailed analysis of the CME features and solar wind parameters, we build a prediction engine taking advantage of 182 previously observed geo-effective partial-/full-halo CMEs and using algorithms of the Support Vector Machine (SVM). We demonstrate that CAT-PUMA is accurate and fast. In particular, predictions after applying CAT-PUMA to a test set, that is unknown to the engine, show a mean absolute prediction error $\sim$5.9 hours of the CME arrival time, with 54\% of the predictions having absolute errors less than 5.9 hours. Comparison with other models reveals that CAT-PUMA has a more accurate prediction for 77\% of the events investigated; and can be carried out very fast, i.e. within minutes after providing the necessary input parameters of a CME. A practical guide containing the CAT-PUMA engine and the source code of two examples are available in the Appendix, allowing the community to perform their own applications for prediction using CAT-PUMA.
\end{abstract}


\section{Introduction} \label{intro}
Coronal Mass Ejections (CMEs) are one of the two major eruptive phenomena (the other ones are flares) occurring within the solar atmosphere with an effect on the heliosphere. CMEs leave the Sun at average speeds around 500 km s$^{-1}$, carry a large amount of magnetized plasma with an average mass $10^{15}$ g into the interplanetary space and carry a huge amount of kinetic energy, often of the order $10^{30}$ erg \citep[for reviews, see e.g.,][and references therein]{Low2000, Chen2011, Webb2012, Gopalswamy2016}. The following observational facts highlight some of the most important aspects why enormous attention has been paid towards CMEs in the past several decades since their first discovery \citep{Hansen1971, Tousey1973}: 1) CMEs are usually accompanied by some other dynamic, large-scale phenomena including e.g. filament eruptions \citep[e.g.,][]{Jing2004, WangY2006, LiuR2010}, flares \citep[e.g.,][]{Harrison1995, Qiu2004, Zhang2012}, magneto-hydrodynamic (MHD) waves \citep[e.g.,][]{Chen2005, Biesecker2012, LiuW2012}, radio bursts \citep[e.g.,][]{Jackson1978, Lantos1981, ShenC2013, ChenY2014} and solar jets \citep[e.g.,][]{ShenY2012, LiuJ2015, ZhengR2016}. Combined studies of CMEs and their accompanying phenomena could improve our understanding of the physical processes taking place in various regimes of the Sun. 2) MHD shocks caused by CMEs could be employed to gain insight into the characteristic properties of the plasma state in the interplanetary space \citep[for reviews, see e.g.,][]{Vrsnak2008}. 3) CMEs occur with a range of rate of abundance both during solar minimum and maximum \citep[e.g.,][]{Gopalswamy2003, Robbrecht2009}, study of which may help us in exploring the solar cycle and dynamo. 4) Shocks and often large amount of magnetic fluxes carried by CMEs could cause severe disturbances in the Earth's magnetosphere \citep[e.g.,][]{WangY2003, WangY2007, ZhangJ2007, Sharma2013, Chi2016} and further affect the operation and working of high-tech facilities like spacecraft, can cause disruption in functioning of modern communication systems (including radio, TV and mobile signals), navigation systems, and affect the working of pipelines and high-voltage power grids.

Besides intensive efforts made towards a better understanding of how CMEs are triggered \citep[e.g.,][]{Gibson1998, Antiochos1999, Lin2000, Forbes2000}, many studies have been focused on predicting the arrival (or transit) times of CMEs at the Earth, having considered their potentials in largely affecting the Earth's magnetosphere and outer atmosphere. This has become one of the most important contents of the so-called space weather forecasting efforts. However, despite of the lack of in-situ observations of the ambient solar wind and CME plasma in the inner heliosphere at CMEs eruption, there are also several further effects that make it more complex and rather challenging to predict CMEs' arrival time, including, e.g., the fact that CMEs may experience significant deflection while traveling in the interplanetary space \citep[e.g.][]{WangY2004, Gui2011, Isavnin2014, Kay2015, Zhuang2017} and that CMEs may interact with each other causing mering or acceleration/deceleration \citep[e.g.,][]{WangY2002, ShenC2012, ShenF2013, Mishra2016, Lugaz2017}.

Current models about the prediction of CME arrival time may be classified into three types: empirical, drag-based and physics-based (MHD) models \citep[for a review, see e.g.,][]{Zhao2014}. Most empirical models use a set of observed CMEs to fit a simple relation (linear or parabolic) between observed CME speeds (and/or accelerations) and their transit times in the interplanetary space \citep[e.g.,][]{Vandas1996, WangYM2002, Manoharan2006, Schwenn2005, Xie2006}. \cite{Vrsnak2007} took the ambient solar wind speed into account in their empirical model, but still utilizing linear least-square fitting. The drag-based models (DBMs) have an advantage over the empirical models that DBMs take into account the speed difference between CMEs and their ambient solar wind, which may cause considerable acceleration or deceleration of CMEs \citep[e.g.,][]{Vrsnak2001, Subramanian2012}. On the other hand, DBMs are based on a hydrodynamic (HD) approach and ignore the potentially important role of magnetic field in the interaction between CMEs and solar wind. Finally physics-based (MHD) models \citep[e.g.,][]{Smith1990, Dryer2001, Moon2002, Toth2005, Detman2006, Feng2006, Feng2007, Riley2012, Riley2013} are mostly utilizing (M)HD simulations employing observations as boundary/initial conditions in the models to perform prediction of the transit times of CMEs. Though, considering the complexity and less prediction errors of physics-based (MHD) models, there are a few drawbacks, e.g., they are still highly idealized and may require extensive computational resources in terms of hardware and CPU time \citep[e.g.,][]{Toth2005}. Complex or not, previous predictions give, on average, around 10-hour mean absolute errors on CME arrival times \citep[see review,][]{Zhao2014}. Employing 3D observations from the STEREO spacecraft, \cite{Mays2013} reduced the mean absolute error to $\sim 8.2$ hours predicting the arrival time of 15 CMEs. Again using STEREO observations, but allowing only very short lead times ($\sim 1$ day), \cite{Mostl2014} further enhanced the performance for the arrival times to $\sim 6.1$ hours after applying empirical corrections to their models. A fast and accurate prediction with large lead time, using only one spacecraft, is therefore still much needed.

In this paper, we propose a new approach to modeling the partial-/full-halo CME Arrival Time Prediction Using Machine learning Algorithms (CAT-PUMA). We will divide 182 geo-effective CMEs observed in the past two decades, i.e. from 1996 to 2015, into two sets: namely, for training and for testing purposes, respectively. All inputs will be only observables. Without a priori assumption or underlying physical theory, our method gives a mean absolute prediction error, around as little as 6 hours. Details on data mining are in Sec.~\ref{data}. Overview of the employed machine learning algorithms and the implemented training process are described in Sec.~\ref{optim}. Results and comparison with previous prediction models are discussed in Sec.~\ref{result}. We summarize in Sec.~\ref{summary}. A practical guide on how to perform predictions with CAT-PUMA is presented in Appendix~\ref{guide}.

\section{Data Mining} \label{data}
To build a suitable set of input for the machine learning algorithms, the first step of our data mining is to construct a list of CMEs that have eventually arrived at Earth and have also caused disturbances to the terrestrial magnetic field. Such CMEs are usually called geo-effective CMEs. We defined four different Python crawlers to automatically gather the onset time, which is usually defined as the first appearance in the Field-of-View (FOV) of SOHO LASCO C2 \citep{Brueckner1995}, and the arrival time of CMEs, which represents the arrival time of interplanetary shocks driven by CMEs hereafter, from the following four lists:

\noindent \textbf{1.} The Richardson and Cane List \citep{Richardson2010}. The list is available at \url{http://www.srl.caltech.edu/ACE/ASC/DATA/level3/icmetable2.htm} and contains various parameters, including the average speed, magnetic field, associated DST index of more than 500 Interplanetary CMEs (ICMEs) from 1996 to 2006 and the onset time of their associated CMEs if observed. We discard events with no or ambiguously associated CMEs, and obtain the onset and arrival time of 186 geo-effective CMEs from this list.

\noindent \textbf{2.} List of Full Halo CMEs provide by the Research Group on Solar-TErrestrial Physics (STEP) at University of Science and Technology of China (USTC) \citep{Shen2013}. A Full halo CME is defined when its angular width observed by SOHO LASCO is 360$^\circ$. This list is available at \url{http://space.ustc.edu.cn/dreams/fhcmes/index.php} and provides the 3D direction, angular width, real and projected velocity of 49 CMEs from 2009 to 2012, and the arrival time of their associated shocks if observed. Events without observation of the associated interplanetary shocks are removed. The onset and arrival times of 24 geo-effective CMEs are obtained from this list.

\noindent \textbf{3.} The George Mason University (GMU) CME/ICME List \citep{Hess2017}. This list contains information similar to that of the Richardson and Cane list of 73 geo-effective CMEs and corresponding ICMEs from 2007 to 2017. It is available at \url{http://solar.gmu.edu/heliophysics/index.php/GMU_CME/ICME_List}. We only select ICME events satisfying the following criterion: i) there are associated shocks and ii) multiple CMEs are not involved. After implementing the selection criteria, 38 events are obtained from this list.

\noindent \textbf{4.} The CME Scoreboard developed at the Community Coordinated Modeling Center (CCMC), NASA. It is a website allowing the community to submit and view the actual and predicted arrival time of CMEs from 2013 to the present (\url{https://kauai.ccmc.gsfc.nasa.gov/CMEscoreboard/}). For our analysis, we remove those events that did not interact with the Earth and those that have a ``note''. Event was labeled with a ``note'' because, e.g., that the target CME did not arrive at Earth, or there was some uncertainty in measuring the shock arrival time, or there were multiple CME events. Here, we obtained 134 CME events from this list.

Combining all four lists, we obtain eventually 382 geo-effective CME events via data-mining. However, there are overlaps between these lists. To remove duplicates, we remove one of such pairs if two CMEs have onset times with a difference less than 1 hour. 90 events are therefore removed.

The {\it SOHO LASCO CME Catalog} (\url{https://cdaw.gsfc.nasa.gov/CME_list/}) provides a database of all CMEs observed by SOHO LASCO from 1996 to 2016 \citep{Gopalswamy2009}. Via matching the onset time of CMEs in our list with the onset time of CMEs recorded in the SOHO LASCO CME Catalog, we obtain various parameters of them including the angular width, average speed, acceleration, final speed  in the FOV of LASCO, estimated mass and main position angle (MPA, corresponding to the position angle of the fastest moving part of the CME's leading edge). The location of the source region of full halo CMEs can be obtained from the {\it SOHO/LASCO Halo CME Catalog} (\url{https://cdaw.gsfc.nasa.gov/CME_list/halo/halo.html}). CMEs that have no source-region information in the above catalog are further investigated manually, one-by-one, to determine their source region location. Further, events from our compiled list are further removed if they have: i) angular width less than 90$^\circ$; ii) no available mass estimation; or iii) ambiguous source region location. Finally, two CMEs at 2003-10-29 20:54 UT and 2011-10-27 12:12 UT are also removed because the first one has incorrect velocity and acceleration estimation; and, the second one erupted with more than a dozen CMEs during that day.

Eventually, after applying all the above selection criteria, we obtain a list of 182 events containing geo-effective CMEs from 1996 to 2015, of which 56 are partial-halo CMEs and 126 are halo CMEs. The average speed of these CMEs FOV ranges from 400 km s$^{-1}$ to 1500 km s$^{-1}$ in the LASCO FOV.

\section{Optimization} \label{optim}
One of the most popular machine learning algorithms is the Support Vector Machine algorithm (SVM). It is a set of supervised learning methods for classification, regression and outliers detection. The original SVMs were linear \citep[see the review][]{Smola2004}, though SVMs are also suitable for conducting nonlinear analysis via mapping input parameters into higher dimensional spaces with different kernel functions. An implementation of the SVM has been integrated into the Python {\it scikit-learn} library \citep{Pedregosa2011}, with an open-source access and well-established documentation (\url{http://scikit-learn.org/stable/}). According to the {\it scikit-learn} documentation, major advantages of the SVM are that it is: 1) effective in high dimensional spaces, 2) still effective even if the number of dimensions is greater than the number of samples, and 3) memory efficient. Besides, it is particularly well-suited for small- or medium-sized datasets \citep{Geron2017}.

Recent works utilizing machine learning algorithms have been mainly focused on solar flare prediction, CME productivity and solar feature identification using classification methods \citep[e.g.,][]{Li2007, Qahwaji2007, Ahmed2013, Bobra2015, Bobra2016, Nishizuka2017} or multi-labeling algorithms \citep[e.g.][]{Yang2017}. However, to the best of our knowledge, the SVM regression algorithm which is suitable for a wide range of solar/space physics research such as solar cycle prediction, DST index prediction and active region occurrence prediction has not yet been widely used by the solar/space physics community. Further, no previous study has attempted to employ the SVM regression algorithm in the context of applying it to the prediction of CME arrival time.

\subsection{Brief Re-cap of SVM Regression} \label{SVR}

\begin{figure}[tbh]
\centering
\includegraphics[width=\hsize]{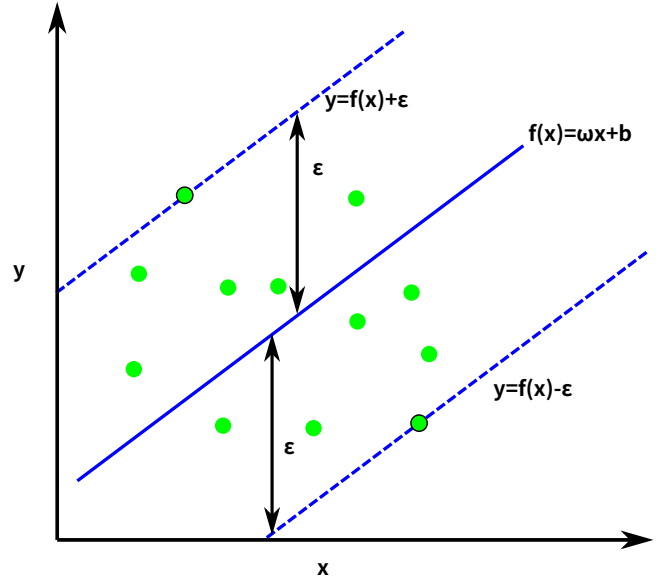}
\caption{An example of the SVM regression in a simple two-dimensional, linear and hard-margin problem. Adopted from Fig. 5-10 in \cite{Geron2017}.}\label{fig:regression}
\end{figure}

To make it simple and clear, we first briefly explain the SVM regression algorithm by demonstrating its capabilities with a simple two-dimensional, linear and hard-margin problem. Let us suppose, there is an input set $x=(x_1, x_2, x_3 ... x_l)$ and a corresponding known result $y=(y_1, y_2, y_3 ... y_l)$ where $l$ is the number of data points. The basic idea of SVM regression is to find a function,

\begin{equation} \label{eq:fx}
f(x) = \omega x + b,
\end{equation}
where, $f(x)$ has at most $\epsilon$ (> 0) deviation from the actual result $y_i$ for all $x_i$ (as shown in Fig.~\ref{fig:regression}). Points at the margins (green dots with black edge) are then called the ``support vectors". New observation $x_{l+1}$ can therefore be taken into Eq.~(\ref{eq:fx}) to yield a prediction for its unknown result $y_{l+1}$.

The solution for the above two-dimensional, linear and hard-margin problem can be extended into multi-dimensional, linear and soft-margin problems. In this case, the target for the SVM regression is to:

\begin{equation} \label{eq:multi}
\begin{aligned}
& \text{minimize} \ \ \ \ \ \frac{1}{2}||\omega{||}^2 + C \sum_1^l{(\xi_i+\xi_i^*)}, \\
& \text{subject to} \ \ \ \begin{cases}
           			y_i - \langle\omega,\ x_i\rangle - b \leq \epsilon + \xi_i, \\
                    \langle\omega,\ x_i\rangle + b - y_i\leq \epsilon + \xi_i, \\
                    \xi_i, \xi_i^* \geq 0,\ i=1,2,3...l,
           		 \end{cases}
\end{aligned}
\end{equation}
where, $x_i=(x_i^1, x_i^2...x_i^n)$ is a $n$-dimensional vector with $n$ the number of features, $i\in[1, l]$, $||\omega||$ is the norm of $\omega$, $\langle\omega,\ x_i\rangle$ is the dot product between $\omega$ and $x_i$, $\xi_i, \xi_i^*$ are the introduced slack variables to perform the feasible constrains for the soft margins \citep{Vapnik2013, Smola2004}. The regularization factor $C > 0$ is introduced to trade-off the amount up to which deviations larger than $\epsilon$ are tolerated. A larger value of $C$ indicates a lower tolerance on errors. 

\begin{figure*}[tbh]
\centering
\includegraphics[width=0.9\hsize]{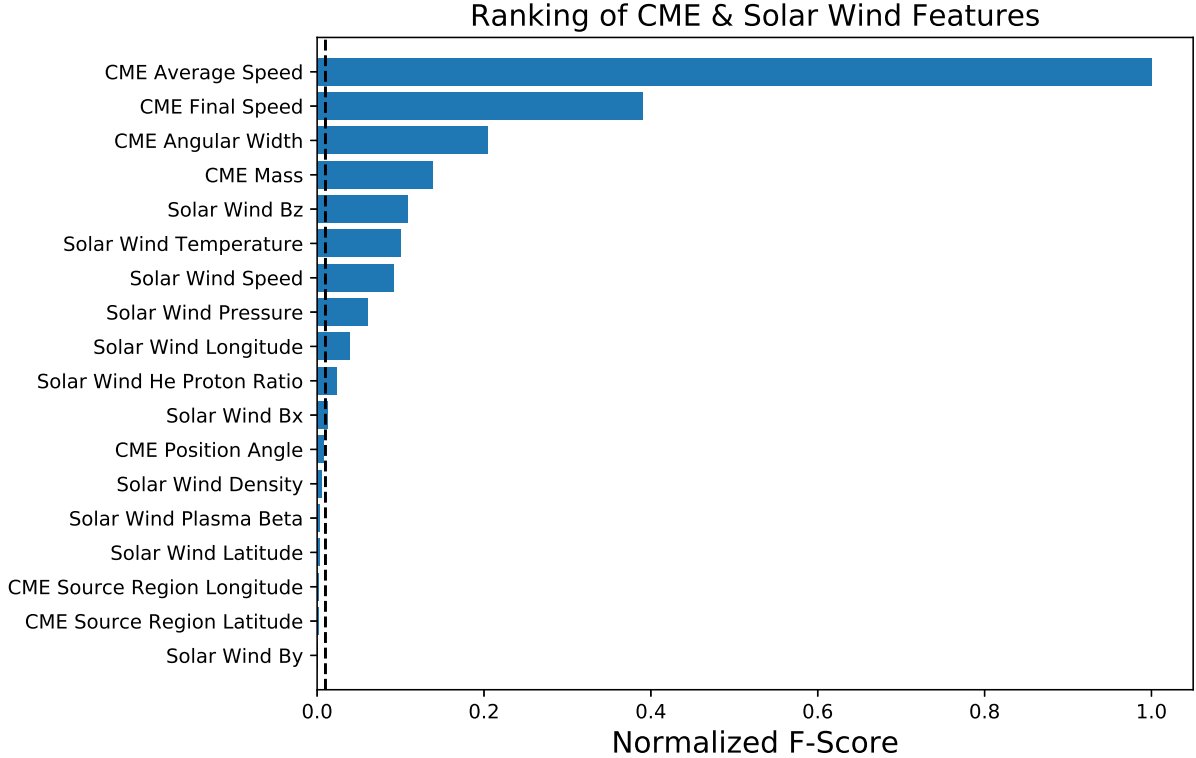}
\caption{Normalized F-scores of all 18 CME and Solar Wind features with $m=6$ hours. The vertical dashed line indicates a normalized F-score of 0.01.}\label{fig:feature}
\end{figure*}

To extend the solution to be suitable for non-linear problems, we map the original non-linear $n$-dimensional input $x$ into a higher-dimensional space $\phi(x)$, in which the problem might be linear. $\phi(x)$ then replaces $x$ in Eq.~\ref{eq:multi}. The most common way to map $x$ into $\phi(x)$ is using kernel functions. One of the most frequently used kernels is the Radial Basis Function (RBF) kernel:

\begin{equation} \label{eq:kernel}
K(x_i, x_j)=\exp{(-\gamma {||x_i - x_j||}^2)},
\end{equation}
where, ${||x_i-x_j||}^2$ is the squared Euclidean distance between the two data points. Here, $\gamma > 0$ defines the area of a single point can influence. A larger $\gamma$ indicates less influence of a point on its neighbors. The description on the SVM regression algorithm above is highly abbreviated. More details can be found in e.g. \cite{Smola2004} and \cite{Vapnik2013}.

Besides $C$ and $\gamma$, another important variable $m$ will be introduced in the rest of this section. The definition of $m$ is given at the beginning of Sec.~\ref{feature}. Processes determining the value of $m$ employed in building the prediction engine are detailed in Sec.~\ref{wind}. Optimization on the selection of parameters $C$ and $\gamma$ are presented in Sec.~\ref{train}.

\subsection{Feature Selection} \label{feature}

Employing the SVM regression algorithms to make predictions of CME arrival time, we take the 182 vectors, of which each contains $n$ parameters of the CME and corresponding solar wind plasma, as $x$ and their actual transit time as $y$. Because, currently it is not feasible to determine the actual background solar wind plasma where a CME is immersed, therefore we use averaged in-situ solar wind parameters at Earth detected from the onset of the CME to $m$ hours later to approximate the actual solar wind parameters at the CME location. In-situ solar wind observations at the Earth, including solar wind $B_x$, $B_y$, $B_z$, plasma density, alpha to proton ratio, flow latitude (North/South direction), flow longitude (East/West direction), plasma beta, pressure, speed and proton temperature, are downloaded from the { \it OMNIWeb Plus} (\url{https://omniweb.gsfc.nasa.gov/}). Together with suitable CME parameters including CME average speed, acceleration, angular width, final speed, mass, MPA, source region latitude and source region longitude described in Sec.~\ref{data}, we have in total 19 ($n=19$) features in the input $x$ space.

However, some of the above features might be important in determining the CME transit time, while some might be irrelevant and unnecessary. Firstly, the CME acceleration is removed from the feature space because it is not independent and basically determined by the CME average speed and final speed. To determine the importance of the rest of the features, following \cite{Bobra2016} but for regression in this case, we use a univariate feature selection tool ({\it sklearn.feature\_selection.SelectKBest}) implemented in the Python {\it scikit-learn} library to test the F-score of every individual feature. For feature $k \in [1, n]$, $x^k$ is a vector with length of $l$. The correlation between $x^k$ and $y$, and the F-score of feature $k$ is then defined as: 

\begin{equation} \label{eq:f-score}
\begin{aligned}
Corr = \frac{(x^k-\overline{x^k})\cdot(y-\overline{y})}{\sigma_{x^k} \sigma_y}, \\
F = \frac{Corr^2}{1-Corr^2} (l-2),
\end{aligned}
\end{equation}
where $l$ is the number of data points as defined in Sec.~\ref{SVR}, $\sigma_{x^k}$ and $\sigma_y$ are the standard deviation of $x^k$ and $y$, respectively. A higher F-score indicates a higher linear correlation between the $k$th feature and the CME transit time $y$ in this case. 

Table~\ref{tb1} lists the rankings of all 18 features (excluding CME acceleration) with $m$ from 1 to $m_{max}$ hours. Again, $m$ represents the number of hours after the onset of the CME. $m_{max}$, the upper limit of $m$, is set as 12 hours after considering the prediction purpose of CAT-PUMA, because an extremely fast CME (with speed over 3000 $km\ s^{-1}$) could reach the Earth within around 13 hours \citep{Gopalswamy2010}. Features with higher F-scores have lower ranking numbers in the table. It turns out that the rankings of all features keep relatively stable. They changes are minor with increasing $m$, especially for the first 12 features in the table. Figure~\ref{fig:feature} depicts the normalized F-scores of all features when $m=6$ hours with the largest F-score as 1.

\begin{table*}
\centering
\caption{Ranking of all 18 features with $m$ from 1 to 12 hours}
\begin{tabular}{c|c|c|c|c|c|>{\bfseries}c|c|c|c|c|c|c}
\hline
Feature & \multicolumn{12}{c}{$m$ (hours)} \\
\cline{2-13}
 & 1 & 2 & 3 & 4 & 5 & 6 & 7 & 8 & 9 & 10 & 11 & 12 \\
\hline
CME Average Speed &  1 &  1 &  1 &  1 &  1 &  1 &  1 &  1 &  1 &  1 &  1 &  1 \\
CME Final Speed &  2 &  2 &  2 &  2 &  2 &  2 &  2 &  2 &  2 &  2 &  2 &  2 \\
CME Angular Width &  3 &  3 &  3 &  3 &  3 &  3 &  3 &  3 &  3 &  3 &  3 &  3 \\
CME Mass &  4 &  4 &  4 &  4 &  4 &  4 &  4 &  5 &  5 &  5 &  4 &  4 \\
Solar Wind B$_z$ &  5 &  5 &  5 &  5 &  5 &  5 &  5 &  4 &  4 &  4 &  5 &  6 \\
Solar Wind Temperature &  7 &  7 &  7 &  7 &  7 &  6 &  6 &  6 &  6 &  6 &  6 &  5 \\
Solar Wind Speed &  6 &  6 &  6 &  6 &  6 &  7 &  7 &  7 &  7 &  7 &  7 &  7 \\
Solar Wind Pressure &  8 &  8 &  8 &  8 &  8 &  8 &  8 &  8 &  8 &  8 &  8 &  8 \\
Solar Wind Longitude & 11 &  9 &  9 &  9 &  9 &  9 &  9 &  9 &  9 &  9 &  9 &  9 \\
CME Acceleration & 10 & 10 & 10 & 10 & 10 & 10 & 10 & 10 & 10 & 10 & 10 & 10 \\
Solar Wind He Proton Ratio & 12 & 12 & 11 & 11 & 11 & 11 & 11 & 11 & 11 & 11 & 11 & 11 \\
Solar Wind B$_x$ &  9 & 11 & 12 & 12 & 12 & 12 & 12 & 12 & 12 & 13 & 15 & 15 \\
CME Position Angle & 13 & 13 & 13 & 13 & 13 & 13 & 13 & 13 & 15 & 14 & 13 & 12 \\
Solar Wind Density & 16 & 17 & 15 & 14 & 14 & 14 & 14 & 14 & 14 & 15 & 14 & 13 \\
Solar Wind Plasma Beta & 19 & 18 & 17 & 15 & 18 & 15 & 15 & 15 & 13 & 12 & 12 & 14 \\
Solar Wind Latitude & 18 & 19 & 19 & 19 & 16 & 16 & 18 & 18 & 17 & 17 & 17 & 16 \\
CME Source Region Longitude & 15 & 15 & 14 & 16 & 15 & 17 & 16 & 16 & 16 & 16 & 16 & 17 \\
CME Source Region Latitude & 17 & 16 & 16 & 18 & 17 & 18 & 17 & 17 & 18 & 18 & 18 & 18 \\
Solar Wind B$_y$ & 14 & 14 & 18 & 17 & 19 & 19 & 19 & 19 & 19 & 19 & 19 & 19 \\
\hline
\end{tabular}
\\
\label{tb1}
Note. The column in bold denotes the ranking of all features at $m=$6 hours, which is the most favorable value in building the prediction engine (Sec.~\ref{wind}).
\end{table*}

Not surprisingly, the average and final CME speeds have the highest F-scores, suggesting their importance in determine the CME transit time. CME angular width and mass rank 3rd and 4th, respectively, which might be due to that the angular width contains information of CME propagating direction; and, CME angular width and mass together imply CME's plasma density which could play an important role in the interaction between the CME and the ambient solar wind. Solar wind features including magnetic field $B_z$ and $B_x$ (strength and poloidal direction of the solar wind magnetic field), proton temperature, plasma pressure, plasma speed, flow longitude (toroidal direction of the solar wind plasma flow) also play important roles with relatively high normalized F-scores. The alpha particle to proton number density ratio in solar wind also ranks high in all the features, which might be caused by that the ratio is usually high in CMEs and Co-rotating Interaction Regions (CIRs) \citep[e.g.,][]{Prise2015}. CMEs/CIRs in front of a CME could potentially influence its transit time. However, this needs to be further examined via analyzing the in-situ observations preceding all the CMEs. Finally, we select 12 features with normalized F-score over 0.01 from high to low as the input of the SVM. CME MPA is also included because it has a normalized Fisher score of 0.008, very close to 0.01.

\subsection{Determine Solar Wind Parameters} \label{wind}

\begin{figure}[tbh]
\centering
\includegraphics[width=\hsize]{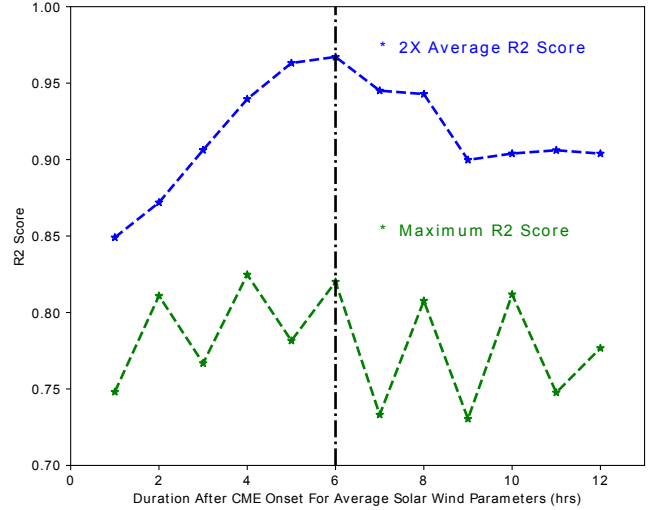}
\caption{Variation of the average (blue curve) and maximum (green curve) $R^2$ scores during a 100000 times training with changing values of $m$ for calculating average solar wind parameters after CME onset.}\label{fig:ip_hour}
\end{figure}

In the previous sub-section, we have shown the result of feature selection using solar wind parameters averaged between the onset time of CMEs and $m$ hours later, where $m$ ranges from 1 to 12. To determine the most favorable value of $m$ in building the prediction engine, 1) we find the optimal $C$ and $\gamma$ for the dataset, followed by 2) training the SVM for 100000 times. 3) we re-calculate the optimal $C$ and $\gamma$ for the best training result. Finally, we repeat the above 3 steps for $m$ ranging from 1 to 12 hours. Details on the first 3 steps will be given in Sec.~\ref{train}. To evaluate how good the models using solar wind parameters with different values of $m$ are, we use the $R^2$ score defined as:

\begin{equation} \label{eq:r2}
R^2 = 1-\frac{\sum\limits_1^l{(y_i-f(x_i))^2}}{\sum\limits_1^l{(y_i-\overline{y})^2}},
\end{equation}
where, $y_i$, $f(x_i)$, $l$ are the same as defined in Sect~\ref{SVR}, and $\overline{y}$ is the average value of $y$. The variation of the maximum and average $R^2$ scores with increasing $m$ is shown in Figure~\ref{fig:ip_hour}. The average $R^2$ score peaks at $m=6$ hours, indicating that the best fitting result is revealed with 6-hour averaged solar wind parameters after CME onset. The maximum $R^2$ score varies ``periodically'' within the range of 0.7 to 0.85 without an overall peak. This ``periodicity'' might have been caused by the combined effect of that 1) 100000 is only a fraction of all $C_{182}^{37}$ ($\sim6\times10^{38}$) possibilities (for further details see Sec.~\ref{train}), thus the best $R^2$ score out of all possibilities cannot always be found during every training, and 2) imperfect stochastic process of the computer in shuffling the dataset (see Paragraph 2, Sec.~\ref{train}). Even though the exact causes of the above ``periodicity'' need further investigation, the variation of the average $R^2$ scores suggests that 100000 is large enough to reflect the overall distribution of the $R^2$ scores.

To summarize the above, we found, using 6-hour averaged solar wind parameters after the CME onset can result in the best output. 

\subsection{Training the SVM} \label{train}

\begin{figure*}[tbh]
\centering
\includegraphics[width=0.9\hsize]{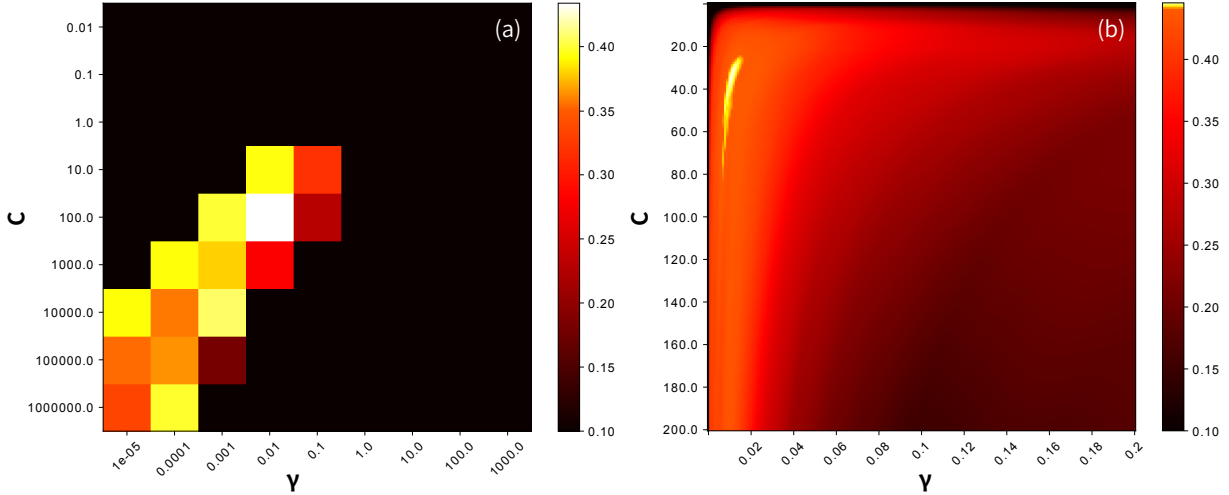}
\caption{Distribution of the average correlation coefficient between the predicted and actual CME transit times of test sets during 3-fold cross-validations repeated for different pairs of $C$ and $\gamma$. In panel (a), $C$ ranges in $[10^{-2}, 10^6]$ and $\gamma$ ranges in $[10^{-5}, 10^3]$. In panel (b), $C$ ranges in $(0, 200]$ and $\gamma$ ranges in $(0, 0.2]$.}\label{fig:cv}
\end{figure*}

One major concern of the SVM regression is the choice of parameters $C$ and $\gamma$. In Sec.~\ref{SVR}, it was demonstrated that the regularization factor $C$ trades off the tolerance on errors. A larger (smaller) $C$ indicates that the SVM will attempt to incorporate more (less) data points. Ill-posed $C$ or $\gamma$ could result in over-fitting (the SVM attempts to fit all data points, which may result in bad prediction for new inputs) or under-fitting (the SVM fits too few data points - it cannot represent the trend of variation of the data). To find the optimal parameters, we utilize the {\it sklearn.model\_selection.GridSearchCV} function to perform exhaustive searches over specified values. First, we build a logarithmic grid with basis of 10, in which $C$ ranges in $[10^{-2}, 10^6]$ and $\gamma$ ranges in $[10^{-5}, 10^3]$, as the input of the {\it GridSearchCV} function. It turns out that the $R^2$ score peaks when $C$ is of the order of $10^2$ and $\gamma$ of $10^{-2}$ (Fig.~\ref{fig:cv}a). Then, we perform the above exhaustive search again but with $C$ in $(0, 200]$ with a step of 1 and $\gamma$ in $(0, 0.2]$ with a step of $10^{-3}$. A more accurate pair of $C$ and $\gamma$ is then found, $C=32$ and $\gamma=0.012$ (Fig.~\ref{fig:cv}b).

For the cross-validation purpose, we split the entire dataset into two subsets: the training set and the test set. \cite{Amari1997} found the optimal number of the test set as $l/\sqrt{2n}$, where $l$ and $n$ are the number of data points and features, respectively. Taking $l=187$ and $n=12$ in our case, we find the partition of the entire dataset between the training set and the test set should be 80\%:20\% (145:37). Using the optimal pair of parameters $C$ and $\gamma$ found above, we feed the training set into the SVM regression algorithm to build a prediction engine. Next, we make a prediction of the CME transit times using the test set and calculate the $R^2$ score between the predicted and actual transit times. To find the best result with the highest $R^2$ score, we randomly shuffle the entire dataset \citep[the order of the events in the dataset is shuffled, which is a general practice to avoid bias, see e.g.,][]{Geron2017} and repeat the above steps (i.e. split the shuffled dataset into the training and test sets, build an engine using the training set and calculate the $R^2$ score of the test set). Theoretically, there are $C_{182}^{37}$ ($\sim6\times10^{38}$) possible combinations of the training set and test set. This is a huge number, and is impossible to exhaustively test all the possibilities given the available computer power for us.

\begin{figure}[tbh]
\centering
\includegraphics[width=\hsize]{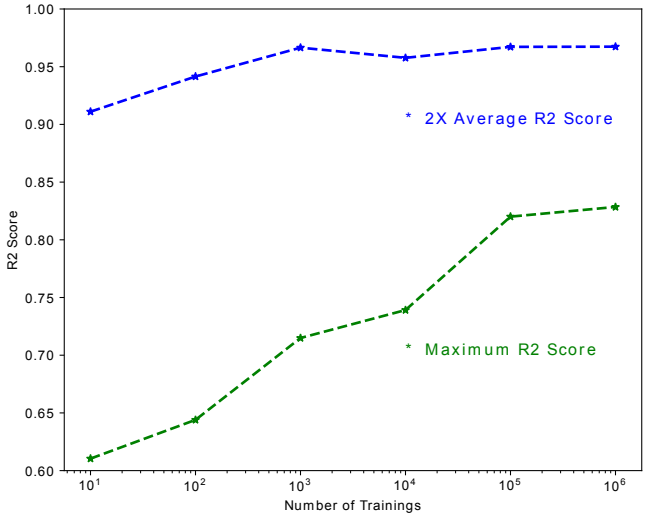}
\caption{Variation of the average (blue curve) and maximum (green curve) $R^2$ scores with increasing number of trainings.}\label{fig:trains}
\end{figure}

Figure.~\ref{fig:trains} shows the variation of the average (blue curve) and maximum (green curve) $R^2$ scores among all the test sets with the increasing number of trainings. The average $R^2$ score increases continuously before the number of trainings reaches 1000, and remains almost unchanged after that. This suggests, when the training is performed over 1000 times, the result can reflect the basic distribution of the $R^2$ scores for all $C_{182}^{37}$ possibilities. The maximum $R^2$ score increases steeply when the number of performed trainings is less than 100000, and yields a similar value when it is increased by a factor of 10. This indicates that it becomes more feasible to find the best engine with increasing number of trainings.

\begin{figure}[tbh]
\centering
\includegraphics[width=0.85\hsize]{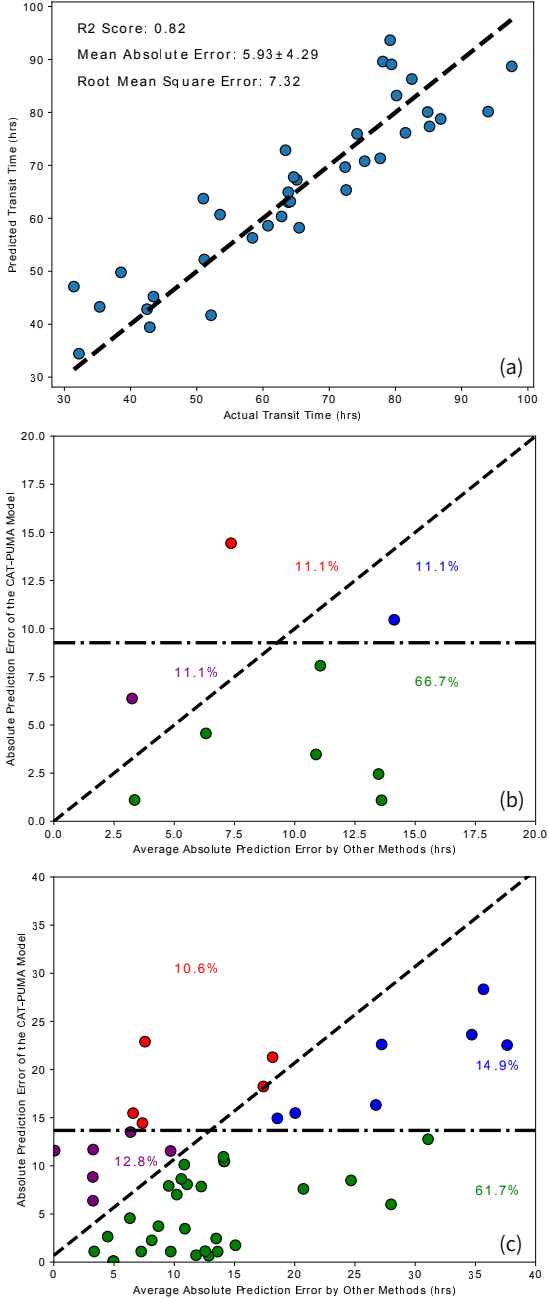}
\caption{(a): Predicted transit time by CAT-PUMA V.S. actual transit time for CMEs in the test set. Black dashed line denotes the same values of the predicted and actual transit time. (b): Comparison between absolute prediction errors by CAT-PUMA and average absolute errors of other methods in the {\it NASA CME Scoreboard}. Only data points included in both the {\it NASA CME Scoreboard} and the test set are shown in this panel. (c): Similar with panel (b), but for all CMEs included in the {\it NASA CME Scoreboard}. Black dashed lines represent that CAT-PUMA has the same prediction errors with the average of other methods. Black dash-dotted lines indicate an absolute error of 9.3 (panel b) and 13.7 (panel c) hours, respectively.}\label{fig:result}
\end{figure}

Considering the above results and reasonable CPU time consumption, we repeat 100000 times of trainings to find the best training set, which results in a highest $R^2$ score of its corresponding test set, to construct the engine. This could be rather costly. However, via paralleling the process employing the open source Message Passing Interface (Open MPI, \url{https://www.open-mpi.org/}), a 100000 times training only takes $\sim$25 minutes on an {\it Intel(R) Core(TM) i7-7770K} desktop with 8 threads. However, we should notice that, training the SVM regression 100000 times cannot always reveal the best result (as shown by the green dashed line in Fig.~\ref{fig:ip_hour}), because 100000 is only a fraction of all the possibilities ($C_{182}^{37}$). Multiple runs are sometimes needed to repeat the 100000 times of trainings.

\section{Results and Comparison} \label{result}

Let us now use the shuffled dataset that yields the highest $R^2$ score of the test set among all the training instances as the input to the engine. The optimal $C=$71 and $\gamma=$0.012 are obtained, again, based on the selected shuffled dataset. Then, we split this dataset into a training set and a test set. CAT-PUMA is then built based on the training set and optimal parameters.

Figure~\ref{fig:result}a shows the relation between the actual transit time and predicted transit time given by CAT-PUMA of the test set. Different blue dots represent different CME events. The black dashed line represents a perfect prediction when the predicted transit time has the same value as the actual transit time. From the distribution of the dots, one sees that they scatter close to the dashed line. The $R^2$ score is $\sim$0.82. The mean absolute error of the prediction is $5.9\pm4.3$ hours, and the root mean square error is 7.3 hours. The probability of detection (POD) is defined as:
\begin{equation}
POD = \frac{Hits}{Hits+Misses}.
\end{equation}
Where, events with absolute prediction errors less and more than 5.9 hours are defined as ``hits'' and ``misses'', respectively. There are 20 events in the test set having absolute prediction errors less than 5.9 hours (Table~\ref{tb2}), giving a POD of 54\%.

\begin{table}
\centering
\caption{Number and percentage of hits and misses in the test set}
\begin{tabular}{c|c|c}
\hline
 & Hits & Misses \\
\hline
Number & 20 & 17 \\
\hline
Percentage & 54\% & 46\% \\
\hline
\end{tabular}
\\
\label{tb2}
\end{table}

There are currently more than a dozen different methods submitted to the {\it NASA CME Scoreboard} by a number of teams to present their predictions of CME arrival times. These methods include empirical, drag-based and physics-based models. More details on the utilized models can be found in the {\it NASA CME Scoreboard} website (\url{https://kauai.ccmc.gsfc.nasa.gov/CMEscoreboard/}) and references therein. Let us now compare the absolute prediction error of CAT-PUMA and the average absolute errors of all other methods available from the {\it NASA CME Scoreboard}, and determine how much progress we have made over the average level of current predictions. Figure~\ref{fig:result}b shows the comparison for CMEs included in both the test set and the {\it NASA CME Scoreboard}, with Figure~\ref{fig:result}c for all CMEs included in the {\it NASA CME Scoreboard}. The dashed lines in both panels indicate when CAT-PUMA has the same prediction error as the average of other models. The dash-dotted lines represent a prediction error level of 9.3 (panel b) and 13.7 (panel c) hours, which are the mean values of the average absolute errors of other methods. Both panels show very similar results. Considering there are only 9 data points in panel (b), we focus on results revealed by panel (c). Green dots (61.7\%) are events of which CAT-PUMA performs better and has errors less than 13.7 hours. Blue dots (14.9\%) are events of which CAT-PUMA performs better but has errors larger than 13.7 hours. Purple dots (12.8\%) are events of which CAT-PUMA performs worse but has errors less than 13.7 hours. Finally, red dots (10.6\%) are events of which CAT-PUMA performs worse and has errors larger than 13.7 hours. In total, CAT-PUMA gives a better prediction for 77\% of the events, and has an error less than 13.7 hours for 74\% of the events.

\section{Summary} \label{summary}
In this paper, we proposed a new tool for partial-/full-halo CME Arrival Time Prediction Using Machine learning Algorithms (CAT-PUMA). During building the prediction engine, we investigated which observed features may be important in determining the CME arrival time via a feature selection process. CME properties including the average speed, final speed, angular width and mass were found to play the most relevant roles in determining the transit time in the interplanetary space. Solar wind parameters including magnetic field $B_z$ and $B_x$, proton temperature, flow speed, flow pressure, flow longitude and alpha particle to proton number density ratio were found important too. 

The average values of solar wind parameters between the onset time of the CME and 6 hours later were found to be the most favorable in building the engine. Considering an average speed of 400 km s$^{-1}$ of the solar wind, it typically takes a 104-hour traveling time from the Sun to Earth. Our results indicate that properties of solar wind detected at Earth might have a periodicity of (104+6)/24=4.6 days. However, this needs to be further examined very carefully by future works.

After obtaining the optimal pair of input parameters $C$ and $\gamma$, the CAT-PUMA engine is then constructed based on the training set that yields a highest F-score of the test set during trainings carried out 100000 times. The constructed engine turns out to have a mean absolute error of about 5.9 hours in predicting the arrival time of CMEs for the test set, with 54\% of the predictions having absolute errors less than 5.9 hours. Comparing with the average performance of other models available in the literature, CAT-PUMA has better predictions in 77\% events and prediction errors less than the mean value of average absolute errors of other models in 74\% events.

To summarize, the main advantages of CAT-PUMA are that: it provides accurate prediction with mean absolute error less than 6 hours; it does not rely on a priori assumption or theory; due to the underlying principles of machine learning, CAT-PUMA can evolve and promisingly improve with more input events in the future; and finally, CAT-PUMA is a very fast open-source tool allowing all interested users to give their own predictions within several minutes after providing necessary inputs. The shortcoming of CAT-PUMA is that it cannot give a prediction whether a CME will hit the Earth or not. 

CAT-PUMA has not included information on the 3D propagating direction of CMEs. We propose that future efforts towards including the 3D propagation direction and 3D de-projected speed, employing either the graduated cylindrical shell (GCS) model with multi-instrument observations \citep{Thernisien2006} or the integrated CME-arrival forecasting (iCAF) system \citep{Zhuang2017}, together with more observed geo-effective CME events, will further improve the prediction accuracy of CAT-PUMA. 

\acknowledgments
\noindent \textbf{Acknowledgements}.
The {\it SOHO LASCO CME catalog} is generated and maintained at the CDAW Data Center by NASA and The Catholic University of America in cooperation with the Naval Research Laboratory. SOHO is a project of international cooperation between ESA and NASA. JL appreciates discussions with Dr. Xin Huang (National Astronomical Observatories, Chinese Academy of Sciences). We thank Dr. Manolis K. Georgoulis (Research Center for Astronomy and Applied Mathematics, Academy of Athens) for his useful advice in improving this paper. JL and RE acknowledge the support (grant number ST/M000826/1) received by the Science and Technology Facility Council (STFC), UK. RE is grateful for the support received from the Royal Society (UK). YW is supported by the grants 41574165 and 41774178 from NSFC.

\appendix
\section{A Practical Guide of Using the CAT-PUMA to Predict CME Arrival Time} \label{guide}
CAT-PUMA is designed to have a very easy user-friendly approach. Users can download the CAT-PUMA engine (``engine.obj"), the source code (``cat\_puma.py") of an example demonstrating how we perform the prediction, and the source code (``cat\_puma\_qt.py'') of a well-designed User Interface (UI) from the following link: \url{https://github.com/PyDL/cat-puma}. All codes are written in Python, and have been tested with Python 2.7 on two {\it Debian}-based x86-64 Linux systems ({\it Ubuntu} and {\it Deepin}) and the x86-64 {\it Windows} 10 system. Modifications of the code will be needed if one prefers to run CAT-PUMA with Python 3. Python libraries, including $\it datetime,\ numpy,\ pandas,\ pickle\ and\ scikit-learn\ (v0.19.1)$, are needed for a proper run of ``cat\_puma.py''. In the following, we first explain the example code ``cat-puma.py" in details.

The first 134 lines in the code import necessary libraries and define functions that will be used in the main program. Lines 138 to 152 define that features we are going to use, value of $m$ (see Sec.~\ref{wind}) and the location of the engine file. Users are not suggested to revise these lines. Lines 155 to 163 are as following:
\begin{verbatim}
# CME Parameters
time = '2015-12-28T12:12:00'  # CME Onset time in LASCO C2
width = 360.  # angular width, degree, set as 360 if it is halo
speed = 1212.  # linear speed in LASCO FOV, km/s
final_speed = 1243.  # second order final speed leaving LASCO FOV, km/s
mass = 1.9e16  # estimated mass using `cme_mass.pro' in SSWIDL or 
               # obtained from the SOHO LASCO CME Catalog
mpa = 163.  # degree, position angle corresponding to the fasted front
actual = '2015-12-31T00:02:00'  # Actual arrival time, set to None if unknown
\end{verbatim}

The above lines define the onset time, angular width, average speed, final speed, estimated mass and MPA of the target CME. These parameters can easily be obtained from the {\it SOHO LASCO CME Catalog} (\url{https://cdaw.gsfc.nasa.gov/CME_list/}) if available or via analyzing LASCO fits files otherwise. Here, we employ a fast halo CME that erupted at 2015-12-28T12:12 UT as the first example. This event was not included in our input dataset when constructing CAT-PUMA. Line 166 defines whether a user prefers to obtain the solar wind parameters automatically. If yes, the code will download solar wind parameters for the specified CME automatically from the {\it OMNIWeb\ Plus} website (\url{https://omniweb.gsfc.nasa.gov/}).

Next, one can then run the code, typically via typing in the command {\it python2 cat\_puma.py}, after following the above instructions to setup the user's own target CME. The prediction will be given within minutes. The prediction result for the above CME is as following (information in the last two lines will not be given if one has not specified the actual arrival time):

\begin{verbatim}
CME with onset time 2015-12-28T12:12:00 UT
will hit the Earth at 2015-12-30T18:29:33 UT
with a transit time of  54.3 hours
The actual arrival time is 2015-12-31T00:02:00 UT
The prediction error is  -5.5 hours
\end{verbatim}

\begin{figure}[tbh]
\centering
\includegraphics[width=0.9\hsize]{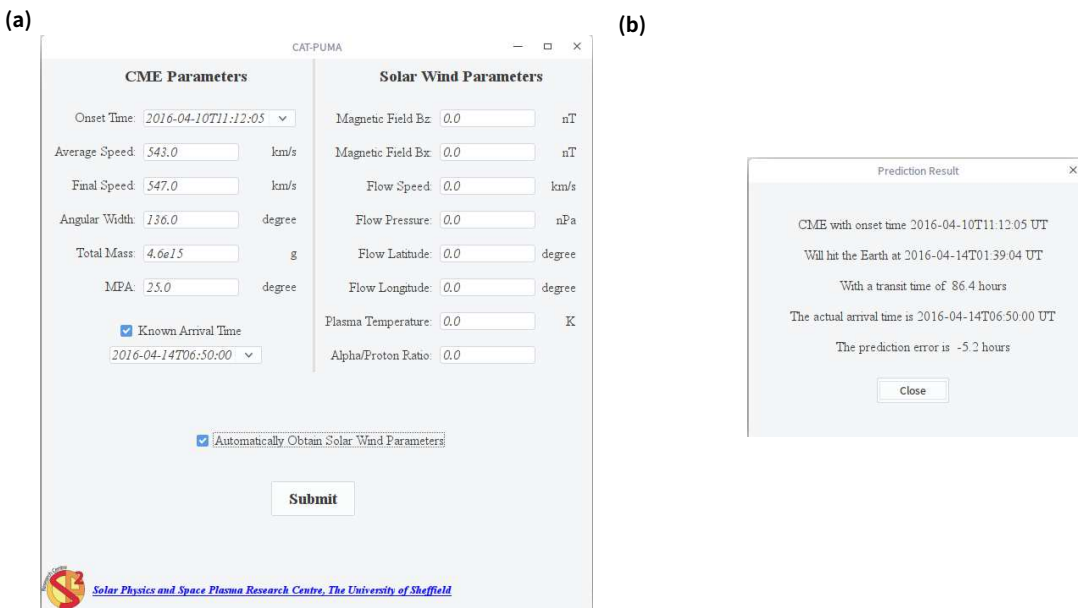}
\caption{The User Interface of CAT-PUMA.}\label{fig:ui}
\end{figure}

Alternatively, one can use the well-designed UI via running the command {\it python2 cat\_puma\_qt.py}. A proper run of it needs additional Python library {\it PyQt5} installed. Let us illustrate how this UI can be used with another example CME that erupted at 2016-04-10T11:12 UT. Again, this event was not included in our input dataset when constructing CAT-PUMA either. Figure~\ref{fig:ui}a shows the UI and corresponding CME parameters for this event. Average speed (543 km s$^{-1}$), final speed (547 km s$^{-1}$), angular width (136$^\circ$) and the MPA (25$^\circ$) were obtained from the {\it SOHO LASCO CME Catalog}. The mass of the CME was estimated by the built-in function ``cme\_mass.pro" in the {\it SolarSoft IDL}. It turns out to be $\sim 4.6\times10^{15}$ g. By checking the option ``Automatically Obtain Solar Wind Parameters'', solar wind parameters are obtained automatically from the {\it OMNIWeb\ Plus} website (\url{https://omniweb.gsfc.nasa.gov/}) after clicking the ``Submit'' button. Then, actual values of the solar wind parameters is shown. Parameters that are not available from the {\it OMNIWeb\ Plus} website are set to 0.00001 (manually input of these parameters are then needed in this case, near real-time solar wind data can be download from the {\it CDAWeb} website \url{https://cdaweb.sci.gsfc.nasa.gov/istp_public/}). Figure~\ref{fig:ui}b shows the prediction result for the above CME, revealing an error of 5.2 hours.

\bibliographystyle{yahapj}

\end{document}